\documentclass[aps, prd, twocolumn, preprintnumbers, superscriptaddress, nofootinbib, floatfix]{revtex4-1}

\usepackage{amsmath}
\usepackage{bm}
\usepackage{amsfonts}
\usepackage{graphicx}
\usepackage{epstopdf}
\usepackage{hyperref}
\usepackage{array}
\usepackage{soul}
\usepackage{color}
\usepackage[utf8]{inputenc}

\enlargethispage{\baselineskip}

\hypersetup{
     colorlinks   = true,
     citecolor    = blue
}

\begin{document}

\title{Hunting for extra dimensions in the shadow of M87*}

\newcommand{\OKCAFF}{\affiliation{The Oskar Klein Centre for Cosmoparticle Physics, Department of Physics, Stockholm University, \\AlbaNova Universitetscentrum, Roslagstullsbacken 21A, SE-106 91 Stockholm, Sweden}}
\newcommand{\FIRSTAFF}{\affiliation{Department of Physics and Astronomy, Uppsala University, L\"agerhyddsv\"agen 1, 75120 Uppsala, Sweden}}
\newcommand{\SECONDAFF}{\affiliation{Nordita, KTH Royal Institute of Technology and Stockholm University, Roslagstullsbacken 23, 10691 Stockholm, Sweden}}
\newcommand{\THIRDAFF}{\affiliation{Kavli Institute for Cosmology (KICC) and Institute of Astronomy,\\University of Cambridge, Madingley Road, Cambridge CB3 0HA, United Kingdom}}
\newcommand{\FOURTHAFF}{\affiliation{Gravitation Astroparticle Physics Amsterdam (GRAPPA), Institute for Theoretical Physics Amsterdam and\\Delta Institute for Theoretical Physics, University of Amsterdam, Science Park 904, 1098 XH Amsterdam, The Netherlands}}
\author{Sunny Vagnozzi}\email{sunny.vagnozzi@fysik.su.se} \OKCAFF \SECONDAFF \THIRDAFF
\author{Luca Visinelli}\email{luca.visinelli@physics.uu.se} \SECONDAFF \FIRSTAFF \FOURTHAFF
\date{\today}

\begin{abstract}
The Event Horizon Telescope has recently provided the first image of the dark shadow around the supermassive black hole M87*. The observation of a highly circular shadow provides strong limits on deviations of M87*'s quadrupole moment from the Kerr value. We show that the absence of such a deviation can be used to constrain the physics of extra dimensions of spacetime. Focusing on the Randall-Sundrum AdS$_5$ brane-world scenario, we show that the observation of M87*'s dark shadow sets the limit $\ell \lesssim 170\,{\rm AU}$, where $\ell$ is the AdS$_5$ curvature radius. This limit is among the first quantitative constraints on exotic physics obtained from the extraordinary first ever image of the dark shadow of a black hole.
\end{abstract}

\maketitle

\section{Introduction}
\label{sec:intro}

The possible existence of extra dimensions, first postulated by Kaluza~\cite{Kaluza:1921tu} and Klein~\cite{Klein:1926fj} in an attempt to unify gravity and electromagnetism, is among the most remarkable predictions of string theory and of $p$-branes~\cite{Dienes:1996du,Green:2012oqa}, albeit virtually inaccessible to current experimental tests.~\footnote{A related important testable prediction of string theory is the existence of a multitude of light pseudoscalar fields known as the axiverse (see e.g.~\cite{Svrcek:2006yi,Arvanitaki:2009fg,Hlozek:2014lca, Marsh:2015xka,Visinelli:2018utg} for works examining the observational testability of this prediction, and also~\cite{Kinney:2018nny} for further related work).} An alternative possibility which can be experimentally probed consists of a non-compactified, large extra dimension: such a scenario occurs for instance in the Randall-Sundrum (RS) model~\cite{Randall:1999ee,Randall:1999vf}. The RS scenario features a five-dimensional AdS space (AdS$_5$) brane-world model where the extra dimension has an infinite size and a negative bulk cosmological constant. The matter fields and the gauge fields of the electromagnetic, weak and strong forces are restricted to a three-dimensional brane, while gravity is free to propagate on the bulk of the AdS$_5$ spacetime. The tension on the brane is chosen so that General Relativity is recovered in the low-energy regime~\cite{Binetruy:1999ut,Binetruy:1999hy}. A review of brane-world gravity models is given in~\cite{Maartens:2010ar}.

In general, the existence of brane-world extra dimensions can be observationally tested through a variety of methods, which include astrophysical observations, precision tests of gravity on small scales, and collider searches~\cite{Antoniadis:1998ig,ArkaniHamed:1998nn,Fitzpatrick:2007qr,Johannsen:2008aa,
Salumbides:2015qwa,Gangopadhyay:2016qqa,Pardo:2018ipy,Chakraborty:2016lxo,Chakravarti:2019aup,Banerjee:2019sae}. In particular, these observations allow to constrain the AdS$_5$ radius of curvature $\ell$, which is a new parameter entering the metric of the space-time. Currently, the best limits on $\ell$ come from the lack of detection of macroscopic forces at laboratory distances, which set the constraint $\ell \lesssim 1\,$mm~\cite{Long:2002wn}. Since gravity and electromagnetism behave in a fundamentally different way in the RS model, studying the propagation of gravitational and electromagnetic signals from a common source is a powerful way to test the existence of brane-world extra dimensions. In a previous paper~\cite{Visinelli:2017bny}, we used the multi-messenger detection of the gravitational wave (GW) event GW170817 and its electromagnetic counterpart GRB170817A~\cite{Monitor:2017mdv,TheLIGOScientific:2017qsa} from a binary neutron star merger to test the RS scenario. The sole measurement of the time lag between the two multi-messenger counterparts allowed us to place the bound $\ell \lesssim {\cal O}({\rm Mpc})$.

Besides altering the propagation of GWs, brane-world extra dimensions are expected to affect the Newtonian potential generated by a massive object lying on the brane (see e.g.~\cite{Randall:1999ee,Randall:1999vf,Callin:2004py,Callin:2004bm,Callin:2004zm}). These modifications can be quite relevant for the metric describing the vicinity of black holes (BHs), since they would alter the BH quadrupole moment with respect to the expected value in the absence of extra dimensions. These modifications are particularly important since the no-hair theorem/conjecture states that BH solutions to the Einstein-Maxwell equations of GR and electromagnetism should be completely characterised by three parameters: mass, electric charge, and angular momentum~\cite{Israel:1967wq,Israel:1967za,Carter:1971zc, Chrusciel:2012jk,Misner:1974qy}. In particular, knowledge of these three quantities would be enough to uniquely determine the BH quadrupole moment: any deviation from such canonical value would be an indication that new physics is at play~\cite{Johannsen:2010xs,Johannsen:2010ru}.

In~\cite{Luminet:1979nyg}, it was shown that the combination of a BH event horizon and gravitational lensing of nearby photons leads to the appearance of a dark shadow, which should be detectable using very long baseline interferometry (VLBI)~\cite{Falcke:1999pj} (see also~\cite{Gralla:2019xty}). Recently, the Event Horizon Telescope (EHT) collaboration~\cite{Fish:2016jil} succeeded for the first time in providing a breathtaking image of the dark shadow of the supermassive black hole (SMBH) M87*, which resides at the center of the giant elliptical galaxy M87~\cite{Akiyama:2019cqa} (see also~\cite{Akiyama:2019brx,Akiyama:2019sww,Akiyama:2019bqs,Akiyama:2019fyp,Akiyama:2019eap}).

In the simplest scenario wherein M87* is a Kerr BH, and gravity behaves as per standard expectations, the dark shadow of M87* should be highly circular. This expectation is indeed met in the results of the EHT collaboration~\cite{Akiyama:2019cqa}, which reports deviations from circularity in the image of the dark shadow of M87* to be of order $10\%$ at most. Still in~\cite{Akiyama:2019cqa}, such a constraint was used to set an upper limit on the relative deviation of M87*'s quadrupole moment fromthe standard value $Q_{\rm Kerr}$: $|\Delta Q/Q_{\rm Kerr}| \equiv \epsilon \lesssim 4$. 

In this short paper, it is our goal to revisit the physics of brane-world extra dimensions in light of the highly circular dark shadow of M87* detected by the EHT collaboration. The rest of this paper is then organized as follows. In Sec.~\ref{sec:rotating}, we first revisit the computation of the quadrupole moment of Kerr BHs within the RS scenario, wherein a fifth dimension with curvature radius $\ell$ is present. Still in Sec.~\ref{sec:rotating}, we then estimate the relative deviation of the quadrupole moment compared to standard expectations as a function of $\ell$, $\epsilon(\ell)$. In Sec.~\ref{sec:results} we use the limit $\epsilon(\ell) \lesssim 4$ set by the EHT collaboration to set a new independent upper limit on the AdS$_5$ curvature radius $\ell$ within the Randall-Sundrum scenario. We provide closing remarks in Sec.~\ref{sec:conclusions}.

\section{Rotating black holes in 4 and 5 dimensions}
\label{sec:rotating}

Let us first consider a Kerr BH in 4 dimensions. The BH has mass $M$ and angular momentum ${\bf J}$, and we further define the rotational parameter $a \equiv |{\bf J}/M|$. Within the usual spherical coordinate system $(r, \theta, \varphi)$, with $r=\sqrt{x^2+y^2+z^2}$, where $(x,y,z)$ are the usual Cartesian coordinates, and in units where $c=G=1$, the square infinitesimal line element of a Kerr BH in 4 dimensions is given by~\cite{Kerr:1963ud}:
\begin{eqnarray}
ds^2 = &-& \left ( 1-\frac{2Mr}{\Sigma} \right )dt^2 + \frac{\Sigma}{\Delta}dr^2+\Sigma d\theta^2 \nonumber \\
&+& \left ( r^2+a^2+\frac{2Mra^2}{\Sigma}\sin^2\theta \right )\sin^2\theta d\varphi^2 \nonumber \\
&-& \frac{4Mra\sin^2\theta}{\Sigma}dtd\varphi\,.
\label{eq:kerr}
\end{eqnarray}
Let us further define the function $\gamma(r,z)$, defined from the real and positive solution of the equation $\gamma^4-(r^2-a^2)\gamma^2-a^2z^2 = 0$. Then, in the weak field limit, the Newtonian gravitational potential $\Phi(r,\theta)$ generated by a source of mass $M$ is given by~\cite{Hernandez:1967zza}:
\begin{eqnarray}
\Phi(r,\theta) = -\frac{M\gamma^3}{\gamma^4+a^2z^2}\,.
\label{eq:potential}
\end{eqnarray}
As shown in~\cite{Hernandez:1967zza}, the Newtonian gravitational potential can then be expanded in multipoles to yield (up to order $(a/r)^2$):
\begin{eqnarray}
\Phi(r,\theta) = -\frac{M}{r} \left [ 1+ \left ( \frac{a}{r} \right )^2{\cal P}_2(\cos\theta) +... \right ]\,,
\label{eq:multipoles}
\end{eqnarray}
where ${\cal P}_2(x) = (3x^2-1)/2$ is the second Legendre polynomial. As expected, for $a=0$ the potential is spherically symmetric, whereas the correction to spherical symmetry is given by the second term on the right-hand side of Eq.~(\ref{eq:multipoles}). In particular, from the correction we read off the Kerr BH quadrupole, given by $Q_{\rm Kerr}=Ma^2$.

We now need to correct Eq.~(\ref{eq:multipoles}) in the presence of a fifth dimension. The process of matter collapse to a BH within the RS model was studied in a series of seminal works~\cite{Argyres:1998qn,Chamblin:1999by,Kraus:1999it,Emparan:1999wa,Garriga:1999yh,Csaki:2000fc,Chamblin:2000md,
Dadhich:2000am,Nojiri:2000yr,Casadio:2000py}. In particular~\cite{Chamblin:1999by} studied the case of a black string living in the AdS$_5$ bulk, whose intersection with the 3-brane would appear as a BH to an observer living on the brane. However, such a ``projected" BH still carries information about the fifth dimension, as discussed in~\cite{Dadhich:2000am}. When projected onto the brane, gravitational field effects in the bulk effectively endow the BH with a \textit{tidal charge} $\beta$, carrying information about the extra dimension and in particular about the AdS$_5$ curvature radius $\ell$. In other words, the BH living on the brane carries an effective charge (which might be positive or negative) due to the tidal influence of the bulk.

These earlier works focused on brane-world BHs for non-rotating spacetimes. As shown in~\cite{Randall:1999ee,Randall:1999vf,Giddings:2000mu}, the continuous spectrum of Kaluza-Klein modes in RS models leads to a correction to the gravitational potential generated by a point source which at low energies goes as $\Delta \Phi \approx 2M\ell^2/(3r^3)$. What about rotating spacetimes, \textit{i.e.} brane-world Kerr BHs? The square infinitesimal line element of a Kerr BH in AdS$_5$ is given by~\cite{Aliev:2005bi}:~\footnote{See also~\cite{Modgil:2001hm,daRocha:2005xk,Aliev:2009cg,Amarilla:2011fx,Aliev:2012rj,
Neves:2012it,Eiroa:2017uuq,Banerjee:2019cjk,Schee:2008kz,Stuchlik:2018qyz} for important works on rotating BHs in brane-world models.}
\begin{eqnarray}
ds^2 = &-&\frac{\Delta}{\Sigma} \left (dt-a\sin^2\theta d\phi \right ) ^2 + \Sigma \left ( \frac{dr^2}{\Delta} + d\phi^2 \right ) \nonumber \\
&+& \frac{\sin^2\theta}{\Sigma} \left [ adt - \left ( r^2+a^2 \right )d\phi \right ]^2\,,
\label{eq:kerr5d}
\end{eqnarray}
where we have defined:
\begin{eqnarray}
\Delta \equiv r^2+a^2-2Mr+\beta\,, \quad \Sigma \equiv r^2+a^2\cos^2\theta\,,
\label{eq:deltasigma}
\end{eqnarray}
with $\beta$ the tidal charge of the BH, which records information from the bulk projected onto the brane. The metric in Eq.~(\ref{eq:kerr5d}) may look familiar: in fact, it is exactly equivalent to the Kerr-Newman metric describing a rotating charged BH in 4 dimensions, where this time the charge $Q$ has been replaced by the tidal charge $\beta$ in the definition of $\Delta$ in Eq.~(\ref{eq:deltasigma}). In fact, it is easy to see that for $\beta=0$ one recovers the Kerr metric of Eq.~(\ref{eq:kerr}). This result is totally analogous to the one found in~\cite{Dadhich:2000am}, where it was found that a Schwarzschild (\textit{i.e.} non-rotating, uncharged) brane-world BH is described by a Reissner-Nordstr\"{o}m (\textit{i.e.} non-rotating, charged) metric, with the BH charge replaced by the tidal charge.

One can show that the earlier estimate of $\Delta \Phi \approx 2M\ell^2/(3r^3)$ is valid to order $(a/r)^2$ (\textit{i.e.} the order relevant for computing the quadrupole) even when considering rotating spacetimes. In other words, the quadrupole does not receive corrections depending on the spin of the BH. This can be seen by following the approach of~\cite{daRocha:2005xk}, wherein one obtains that the leading order correction scales as $\ell^2a^2/r^4$, and thus does not affect the computation of the quadrupole. In summary, the Newtonian gravitational potential for a Kerr BH in the RS model is given, up to order $(a/r)^2$, by:
\begin{eqnarray}
\Phi(r,\theta) = -\frac{GM}{r} \left [ 1+ \left ( \frac{a}{r} \right )^2{\cal P}_2(\cos\theta)+\frac{2\ell^2}{3r^2} +... \right ].
\label{eq:multipolesbrane}
\end{eqnarray}

In all previous equations, $\theta$ is the BH observation angle, which corresponds to the angle between the line-of-sight and the spin axis. For the specific case of M87*, the jet inclination with respect to the line-of-sight is $\theta_{\rm jet} \simeq 17^{\circ}$~\cite{Mertens:2016rhi}, which coincides with the angle between the line-of-sight and the spin axis $\theta$ under the well-motivated assumption that the jet is powered by the spin of the BH and is approximately aligned with the spin axis (through the Blandford-Znajek mechanism~\cite{Blandford:1977ds} or variants thereof). Hereafter, we shall therefore set $\theta = \theta_{\rm jet} \simeq 17^{\circ}$. Notice that for such low observation angle, ${\cal P}_2(\cos\theta) \approx 1$, an approximation which for simplicity we adopt moving forward.~\footnote{There are in fact indications that $\theta$ might even be lower than $17^{\circ}$~\cite{Rieger:2012en}. This would make our approximation of ${\cal P}_2(\cos\theta) \approx 1$ even more accurate.}

\section{Results}
\label{sec:results}

Using the modified Newtonian potential in the presence of extra dimensions estimated in Eq.~(\ref{eq:multipolesbrane}), and using our approximation for ${\cal P}_2(\cos\theta) \approx 1$, we can read off the quadrupole moment $Q_{\rm Kerr,RS} \approx Ma^2+2M\ell^2/3$ and correspondingly the deviation from the Kerr value $\Delta Q \equiv Q_{\rm Kerr,RS}-Q_{\rm Kerr}$, with $Q_{\rm Kerr}=Ma^2$. We find that the deviation $\Delta Q$, and correspondingly the relative deviation $\epsilon$, are given by:
\begin{eqnarray}
\Delta Q &\approx& \frac{2M\ell^2}{3}\,, \label{eq:deltaq}\\
\epsilon &\equiv& \frac{\Delta Q}{Q_{\rm Kerr}} \approx \frac{2\ell^2}{3a^2}\,. \label{eq:epsilon}
\end{eqnarray}

Using our estimate for $\Delta Q$ together with the upper limit on $\epsilon$ reported by the EHT collaboration implies:
\begin{eqnarray}
\epsilon \approx \frac{2}{3}\frac{\ell^2}{a^2} \lesssim 4 \implies \frac{2}{3}\ell^2 \lesssim 4a^2\,.
\label{eq:upperlimit}
\end{eqnarray}
However, for a Kerr BH the angular momentum is bounded by the Kerr limit $a \leq M$, which implies that $4a^2 \lesssim R_s^2$, where $R_s=2M$ is the Schwarzschild radius for the object in question. Armed with this further inequality, we see that Eq.~(\ref{eq:upperlimit}) implies:
\begin{eqnarray}
\ell \lesssim \sqrt{\frac{3}{2}}R_s\,.
\label{eq:upperlimitnew}
\end{eqnarray}
In~\cite{Akiyama:2019eap}, the mass of M87* was estimated to be $M_{M87*} = (6.6\pm 0.7)\times10^{9}\,M_{\odot}$, in units of the mass of the Sun $M_{\odot} \approx 2\times 10^{30}\,$kg. Using this result, we find $R_s \approx 2.5\times 10^{13}\,{\rm m}$, which finally results in the following upper limit on AdS$_5$ curvature radius:
\begin{eqnarray}
\ell \lesssim 2.5\times 10^{13}\,{\rm m} \approx 170\,{\rm AU}\,,
\label{eq:upperlimitfinal}
\end{eqnarray}
with $1\,{\rm AU}$ being one astronomical unit.

The upper limit obtained in Eq.~(\ref{eq:upperlimitfinal}) is still not competitive with the ${\cal O}({\rm mm})$ constraint obtained in~\cite{Long:2002wn}, but represents a substantial improvement over the limit we obtained from GW170817 in~\cite{Visinelli:2017bny}. More importantly, it is an independent limit and among the first limits on exotic physics obtained by imaging the dark shadow of M87*.~\footnote{See also~\cite{Giddings:2019jnp,Moffat:2019uxp,Nokhrina:2019sxv,Abdikamalov:2019ztb,Held:2019xde,Wei:2019pjf,Tamburini:2019vrf,Davoudiasl:2019nlo, Ovgun:2019yor,Cunha:2019dwb,Wang:2019tjc,Hui:2019aqm,Bambi:2019tjh,Konoplya:2019sns,Nemmen:2019idv,Chen:2019fsq,Gyulchev:2019tvk,Shaikh:2019jfr, Firouzjaee:2019aij,Konoplya:2019nzp,Kokubu:2019jdx,Giddings:2019vvj,Kawashima:2019ljv,Kumaran:2019qqp,Bar:2019pnz,Jusufi:2019nrn,Exirifard:2019ile,
Roy:2019esk,Bambi:2019xzp,Long:2019nox} for other works exploring fundamental physics and properties of BHs or BH mimickers in light of the imaged dark shadow of M87* or future observations from the EHT.}

\section{Conclusions}
\label{sec:conclusions}

In this short paper we have shown how the image of the dark shadow of M87*, recently provided by the Event Horizon Telescope collaboration, allows us to probe the physics of extra dimensions. Focusing on the Randall-Sundrum brane-world scenario, wherein our Universe consists of a 3-brane within an AdS$_5$ space, we have set an upper limit on the AdS$_5$ curvature radius $\ell$, finding $\ell \lesssim 170\,{\rm AU}$. This upper limit is far from being competitive with current constraints on extra dimensions from precision tests of gravity on ${\cal O}({\rm mm})$ scale. While weak, we believe this constraint is nonetheless valuable because it is obtained in a completely independent way with respect to previous limits. Moreover, it serves as a proof-of-principle for the possibility of constraining fundamental physics already from the first image of the dark shadow of a black hole.

The coming years will certainly be an extremely exciting time as far as tests of General Relativity and alternative theories of gravity are concerned. The success of the Event Horizon Telescope in imaging the dark shadow of M87* pairs up with LIGO's success in detecting an ever-increasing number of GW events: together, the two provide us further tests of the behaviour of gravity on astronomical and cosmological scales. It is exciting to see how we can use available data and build upon the considerable efforts put into action to deploy the first ever image of a black hole shadow, to probe physics describing the very structure of our space-time, testing the possible existence of extra dimensions beyond those which our senses are able to experience. We look forward to improvements in VLBI technology, as well as the delivery of the first image of Sgr A*, which will allow us to refine tests of fundamental physics by imaging the dark shadows of astrophysical black holes.

\begin{acknowledgments}
We thank Cosimo Bambi, Katherine Freese, and Maurizio Giannotti for useful discussions on the subject. We acknowledge support by the Vetenskapsr\r{a}det (Swedish Research Council) through contract No. 638-2013-8993 and the Oskar Klein Centre for Cosmoparticle Physics. S.V. acknowledges support from the Isaac Newton Trust and the Kavli Foundation through a Newton-Kavli fellowship. We thank the University of Michigan, where part of this work was conducted, for hospitality. L.V. thanks Barry University and the University of Florida for hospitality and support.
\end{acknowledgments}

\bibliographystyle{JHEP}
\bibliography{BHImage}

\end{document}